\documentclass{iaus}
\usepackage{graphicx}

\title[Populations of massive stars in galaxies] 
{Populations of massive stars in galaxies, implications
for the stellar evolution theory}

\author[Meynet, Eggenberger, Maeder]   
{Georges Meynet$^1$, Patrick Eggenberger$^2$ \break \and Andr\'e Maeder$^1$}

\affiliation{$^1$Geneva Observatory, University of Geneva, CH-1290 Sauverny, Switzerland \break email: georges.meynet@obs.unige.ch; Andre.Maeder@obs.unige.ch\\[\affilskip]
$^2$Li\`ege University, AGO Department, B-4000 Li\`ege 1 \break email: eggenberger@astro.ulg.ac.be}

\pubyear{2007}
\volume{241}  
\pagerange{NNN--NNN}
\date{?? and in revised form ??}
\jname{Proceedings Title IAU Symposium}
\editors{A.C. Editor, B.D. Editor \& C.E. Editor, eds.}
\begin{document}

\maketitle

\begin{abstract}
After a brief review of the observational evidences indicating
how the populations of Be stars, red/blue supergiants, Wolf-Rayet stars
vary as a function of metallicity, we discuss
the implications of these observed trend for our understanding of the massive
star evolution. We show how the inclusion of the effects of rotation
in stellar models improves significantly the correspondence between theory and
observation.
\end{abstract}

\firstsection 
\section{Massive star populations and stellar models}


For constraining stellar models, the best way is to compare the results of tailored theoretical models with well observed characteristics of single stars. The case of the Sun is in that
respect exemplar. Considering populations of massive stars for constraining stellar models may appear at first sight a loose way to proceed since in addition to the physics of stars, other
ingredients enter into the comparison as the Initial Mass Function (IMF) and the Star Formation History.
However in some circumstances, observed stellar populations may provide powerful
constraints. For instance, IMF has no influence on ratios of massive stars 
involving stars of about the same range of initial masses (as for instance ratios of blue to red supergiants, of Wolf-Rayet (WR) to O-type stars). 
Star formation history is not involved either provided we concentrate
on regions of constant star formation ({\it i.e.} constant during at least the last 10-20 millions years). In that case the number ratios of massive stars in two different evolutionary stages is given
by the ratio of the durations of these evolutionary stages\footnote{Let us note that for starburst the situations is different, because the massive star population varies as a function of the age and
depends also on the duration and intensity of the initial burst of star formation.}. In the following
we briefly recall the main characteristics of the observed populations of Be stars, red and blue supergiants and Wolf-Rayet stars and deduce some consequences for the stellar models. The main
conclusion is that, in all cases, rotation appears as a key physical ingredient in order
to reproduce the mean features of these populations. We end by saying a few words about the
consequences for nucleosynthesis, and again we confirm that rotation plays a key role in this
area also. Before addressing these questions, let us say a few words about the effects of rotation.

\section{Rotating models}

The physics of rotation has been extensively discussed in previous papers (see e.g. Zahn, 1992; Maeder \& Meynet 2001)
to which the reader can refer. Massive star
grids of non-rotating and rotating models have been made  at $Z$ = 0.040, 0.020, 0.008, 0.004 and $10^{-5}$
(Meynet \& Maeder 2005).
The rotating models have an initial velocity $v_{\mathrm{ini}}$ of 300 km/s,
which gives an average velocity of 220 km/s during the MS phase, corresponding
to observations. The main effects of rotation are the following:
\begin{itemize}
\item Rotation increases  the MS lifetime with respect to non--rotating models (up to
 about 40 \% for the most massive stars and for very fast rotation).
\item The values assigned from isochrones with an average rotation velocity  typically lead to ages 25\% larger than without rotation.
\item These models account for the observed changes of the surface abundances for OB main sequence 
and supergiants stars (see also Heger \& Langer 2000).
\item Steeper gradients of internal rotation $\Omega$ are built at lower $Z$.
The steeper $\Omega$--gradient at lower $Z$  favors mixing.
There are 2 reasons for the steeper $\Omega$--gradients. 
One is the higher compactness of the 
star at lower $Z$. The second one is that at lower $Z$, the density of the outer layers
is higher, thus the meridional currents are slowlyer. This produces less outward transport of angular momentum and 
contributes to steepen 
the $\Omega$--gradient.
\item  At lower $Z$, rotating stars more easily reach break--up velocities and may
stay at break-up for a substantial fraction of the MS phase.
\end{itemize}
 
The different scenarios for the evolution of massive stars are indicated below. 

\noindent
{\bf \underline{{$M >90  M_{\odot}$}}}:  O –- Of –- WNL –- (WNE) -– WCL –- WCE -– 
SN (Hypernova low Z  ?)\\
\noindent
{\bf \underline{{$60-90 \; M_{\odot}$}}}: O –- Of/WNL$\Leftrightarrow$LBV -– WNL(H poor)-– WCL-E -– SN(SNIIn?)\\
\noindent
{\bf \underline{{$40-60 \; M_{\odot}$}}}: O –- BSG –-  LBV $\Leftrightarrow$ WNL -–(WNE) -- WCL-E –- SN(SNIb) \\
\hspace*{5.9cm}  - WCL-E - WO – SN (SNIc) \\
\noindent
{\bf \underline{{$30-40 \; M_{\odot}$}}}:  O –- BSG –- RSG  --  WNE –- WCE -– SN(SNIb)\\
\hspace*{4.0cm}                        OH/IR $\Leftrightarrow$ LBV ? \\
\noindent
{\bf \underline{{$25-30 \; M_{\odot}$}}}: O -–(BSG)–-  RSG  -- BSG (blue loop) -- RSG  -- SN(SNIIb, SNIIL)\\
\noindent
{\bf \underline{{$10-25 \; M_{\odot}$}}}: O –-  RSG -– (Cepheid loop, $M < 15 \; M_{\odot}$) – RSG -- 
SN (SNIIL, SNIIp)\\ 

The sign $\Leftrightarrow$ means back and forth motions between the two  stages. The limits 
between the various scenarios  depend on metallicity $Z$ and rotation.
  The various types of supernovae are tentatively indicated.

\section{The Be stars}

Be stars are emission line stars. Emission originates in a circumstellar outflowing disk.
How do these disks form~? How long are their lifetimes~? Are they intermittent~? 
Are they Keplerian~? Many of these questions are still subject of lively debate. A point however which seems well accepted is the fact that
the origin of a disk might be connected to the fast
rotation of the star (Pelupessy et al. 2000).
Martayan et al. (2006) showed that the initial velocities
of the Be stars is significantly higher than the initial velocities of the normal B stars,
giving some support to the view that
only stars with a sufficiently high initial velocity can go through a Be star episode.

\begin{figure}
\includegraphics[width=2in,height=2.1in]{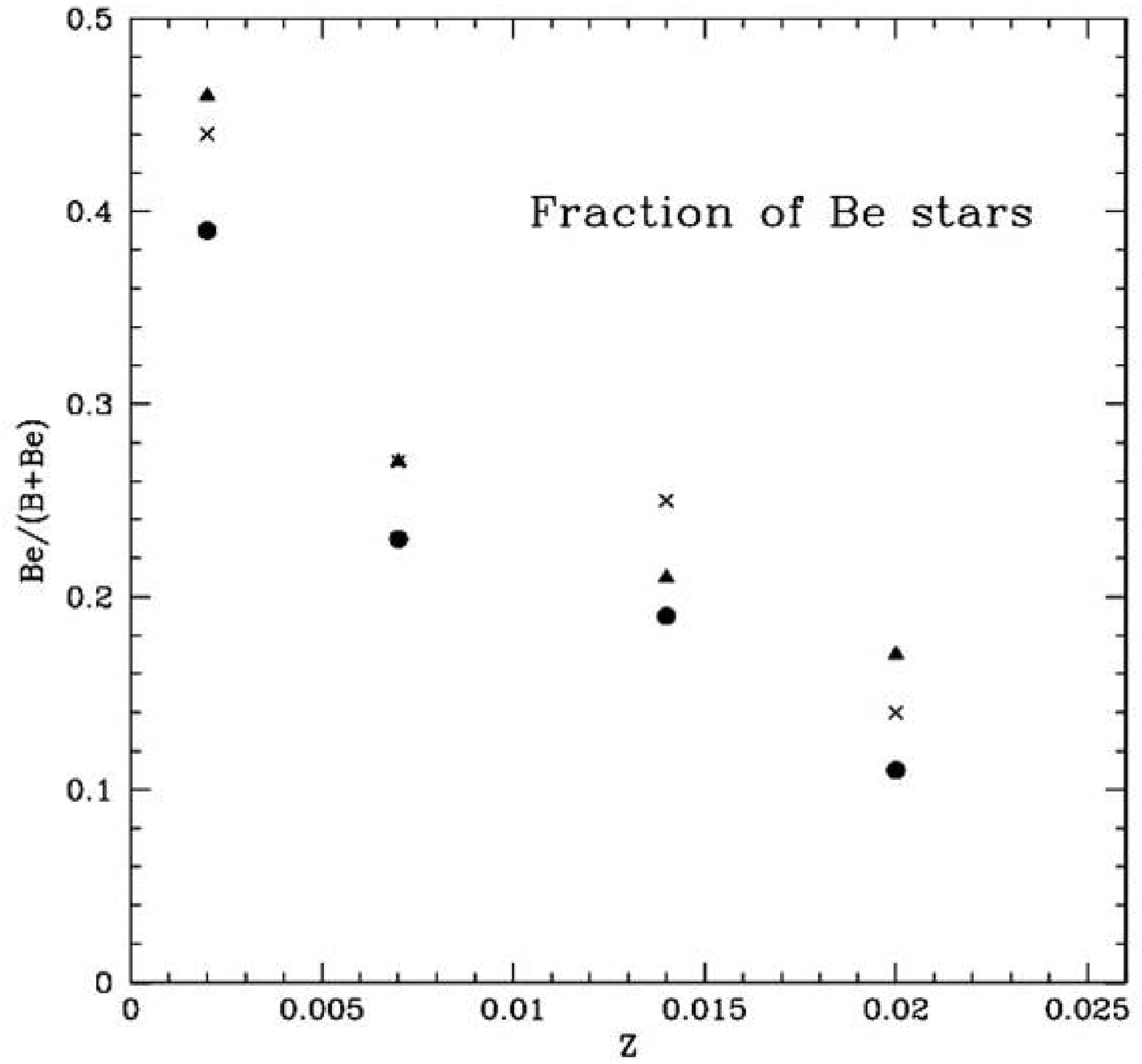}
\hfill
\includegraphics[width=3in,height=2.1in]{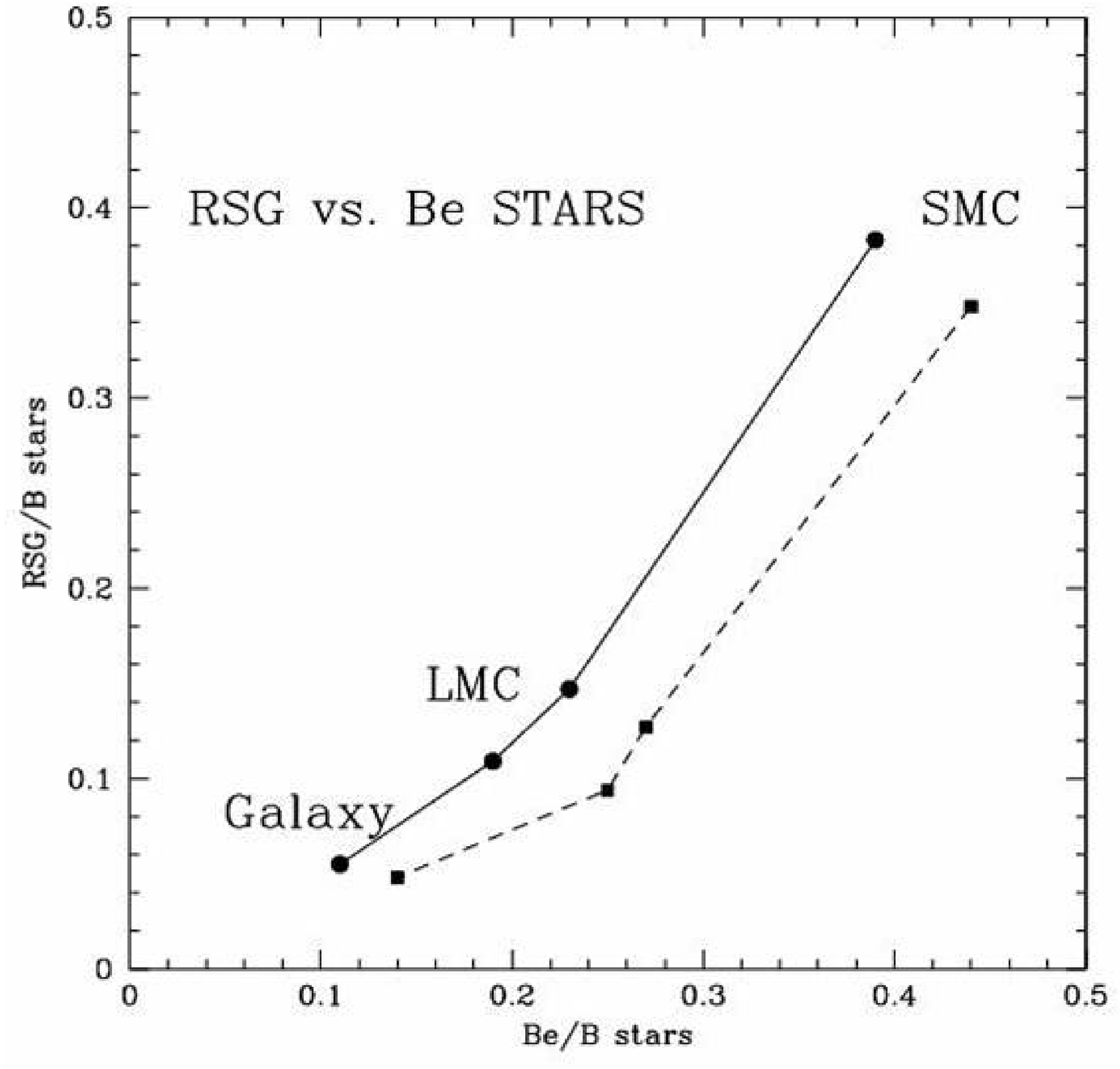}
\caption{{\it Left panel}: Variation as a function of metallicity of the number ratio
of Be stars to B and Be stars (Maeder et al. 1999) 
{\it Right panel}: Correlation between the the number of red supergiants and the number ratio
of Be to B stars.}\label{fig0}
\end{figure}

The population of Be stars varies with the metallicity.
Fig.~\ref{fig0}, {\it left panel}, shows that the number of Be stars with respect to the total number of B stars
(B and Be stars) in cluster with ages between 10 and 25 My (mass at the turn off between
about 9 and 15M$_\odot$) increases with decreasing metallicity (Maeder et al. 1999).
Such a trend has been recently confirmed by Wisniewski \& Bjorkman (2006).
Very interestingly there appears to be a correlation between the frequency of Be stars and that of red supergiants as shown in Fig.~\ref{fig0} {\it right panel}.

These observations indicate that metallicity plays a role in the Be phenomenon and
provides hints on the way surface velocity may evolve differently for stars of different
initial metallicities. The correlation of Be star populations with those of red supergiants
can be seen as an indication that fast rotation not only favors the formation of Be stars
but also of that red supergiants.

As long as the mass losses by stellar winds are not too strong (which is the case for either low metallicity and/or
initial masses below about 15 M$_\odot$), the surface equatorial velocity increases with time during the MS phase. This is due to the coupling between the core and the envelope exerted by the meridional currents. These currents tends to slow down the contracting, still faster spinning core, and to accelerate the expanding outer envelope. Due to this effect, a star beginning its evolution on the ZAMS with a sufficiently high initial rotation rate, may reach the break-up limit at a given point during the MS phase (the break-up limit is the point at which the equatorial velocity is such that the centrifugal acceleration is equal to the gravity at the equator).

\begin{figure}
\includegraphics[width=4.0in,height=3.0in]{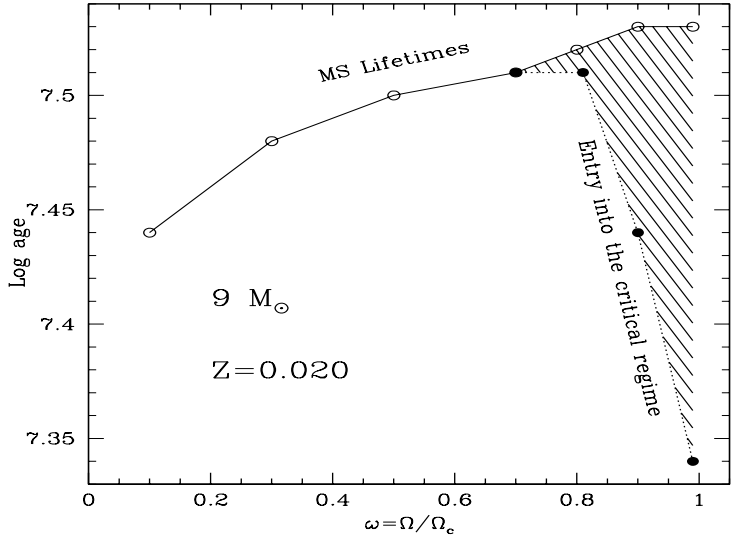}
\hfill
\caption{The MS lifetime of Z=0.02 9M$_\odot$ is plotted for different initial velocities (continuous line with empty circles). The age at which the star encounters for the first time the critical limit is indicated (dotted line with full circles). The hatched zone shows the region where stars at the critical limit are expected.
}\label{fig6}
\end{figure}

In Fig.~\ref{fig6}, the lifetimes on the MS of different 9 M$_\odot$ stellar models
corresponding to various initial values of $\Omega/\Omega_{\rm crit}$ are shown as well
as the time at which the surface equatorial velocity reaches the critical limit.
On can see that
\begin{itemize}
\item Only those stars, beginning with an initial angular velocity on the ZAMS such that it corresponds to about 70\% of the critical angular velocity will reach the critical limit during the MS phase.
\item The model starting with $\Omega/\Omega_{\rm crit}=0.99$
reaches the critical limit only after an age of ~22 millions years. This comes from the fact that, at the very beginning of the MS phase, meridional currents transport 
angular momentum from the outer part of the star to the inner one, accelerating the core and slowing down the envelope. Thus the star first evolves away from the critical limit. This phase is quite short lasting only a few percents of the MS lifetime. Then the meridional currents reverse in the outer layers making the surface velocity to approach the critical limit.
\item Those stars which reach the critical limit, do it at the end of the MS phase.
\end{itemize}
If we accept the premise that a Be star is a star near the break-up limit, then Be stars might naturally arise from the normal evolution of stars beginning their life with a sufficiently high
velocity. 
The question is of course to know if such a scenario might account for the observed ratios
and their observed variation with the metallicity (see Fig.~\ref{fig0}). While these points remain to
be carefully examined, one can already say that from a theoretical point of view, one would expect
a higher proportion of Be stars at low Z if the distribution of the initial velocity is biased toward faster rotators at low $Z$ and/or if, due to the weaker stellar winds at low metallicity, less angular momentum is removed from the surface. 

\section{The supergiants}

\begin{figure}
\includegraphics[width=2.5in,height=2.1in]{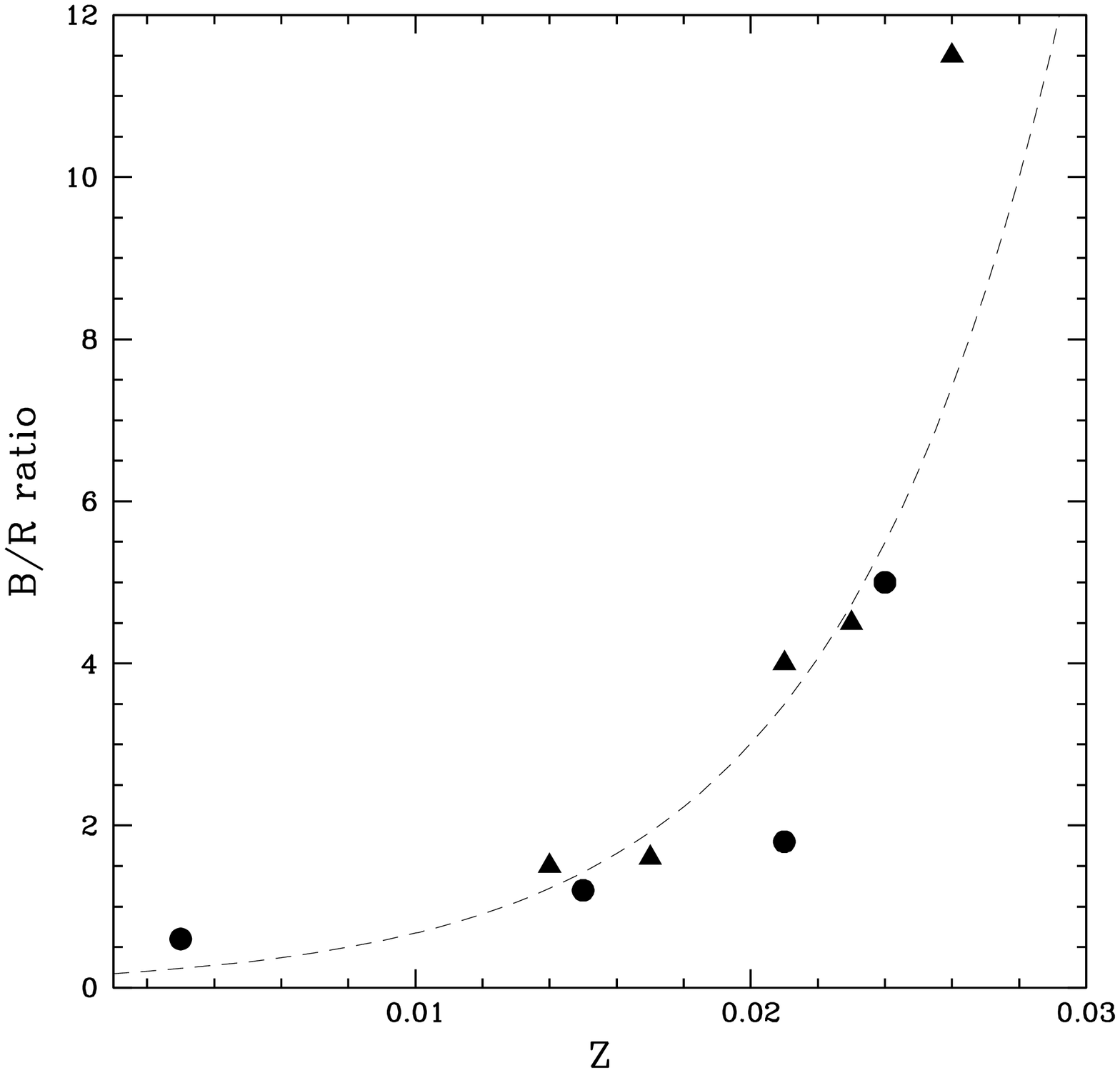}
\hfill
\includegraphics[width=2.5in,height=2.1in]{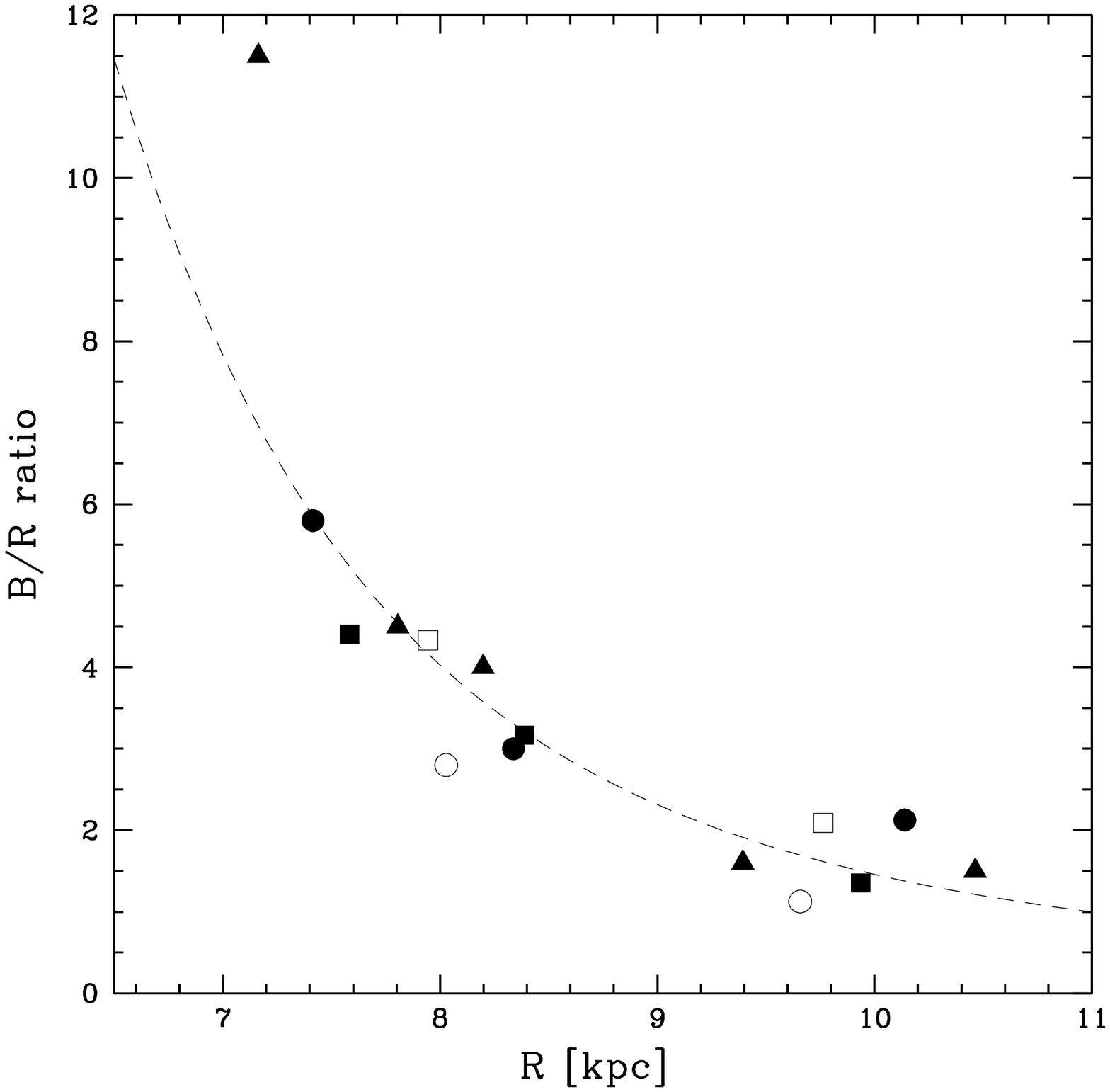}
\caption{{\it Left panel}:  $B/R$ ratio in the Galaxy and the SMC for clusters with $\log age$ between 6.8 and 7.5. The distinction between blue and red
supergiants is based on spectroscopic measurements. The triangles refer to $B$
including O, B and A supergiants.
The dots refer to $B$ including only B supergiants. The dashed curve corresponds to the
fit for $B$ including O, B and A supergiants with $(B/R)_{\odot}=3.0$.
{\it Right panel}: $B/R$ ratio in the Galaxy for different age intervals, with distinction between blue and red
supergiants based on spectroscopic measurements.
$B$ includes O, B and A supergiants. The dashed curve corresponds to the same 
fit as in the left panel. For a detailed description of the figure see Eggenberger et al. (2002)}\label{fig1}
\end{figure}

In Fig.~\ref{fig1}, {\it left panel}, is shown the variation with the metallicity of the number ratio of blue to red supergiants (B/R), and ({\it right panel}) its variation as a function of the galactocentric distance in the Milky Way (Meylan \& Maeder 1982; Eggenberger et al. 2002). Clearly, the B/R ratio increases when the metallicity increases, while standard stellar evolution models ({\it i.e.} models without any extra-mixing processes) predict that this ratio should decrease with increasing metallicity. 
Dohm-Palmer \& Skillman (2002) have
estimated the B/R ratio in the dwarf irregular galaxy Sextans A for various age bins spanning
a range between 20 and 140 My. They find that the observed B/R are lower than those given by standard models by a factor 2, indicating that too few red supergiants are predicted by these models. 

This disagreement implies that no reliable predictions can be made concerning
the nature of the supernova progenitors in different environments, or the populations of supergiants in galaxies. 
The B/R ratio also constitutes an important and sensitive test for stellar evolution models, because it is very sensitive to mass
loss, convection and mixing processes (Langer \& Maeder 1995). 
Thus, the problem of the blue to red supergiant ratio ($B/R$ ratio)
remains one of the most severe problems in stellar evolution.

\begin{figure}
\includegraphics[width=2.5in,height=2.1in]{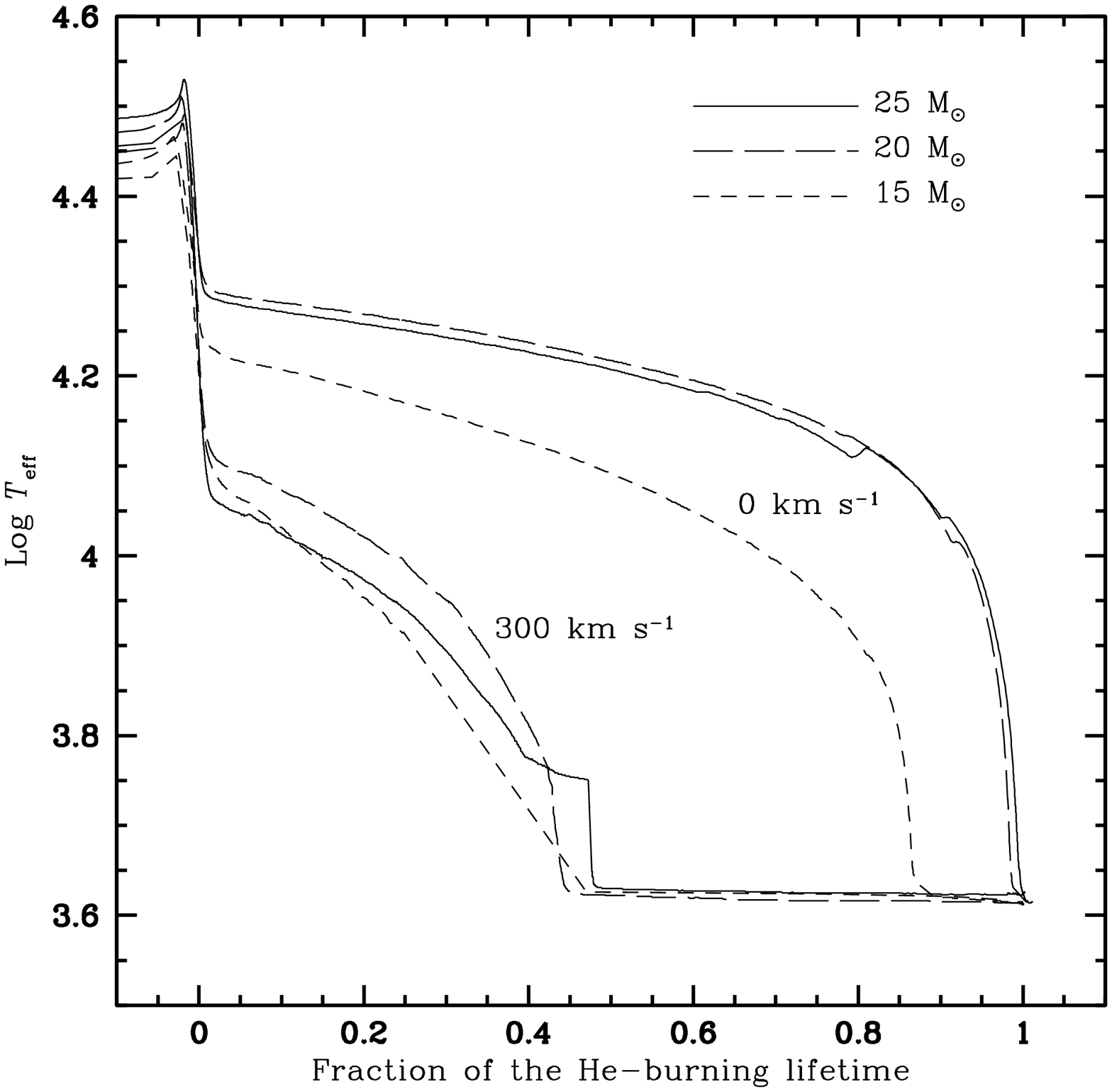}
\hfill
\includegraphics[width=2.5in,height=2.1in]{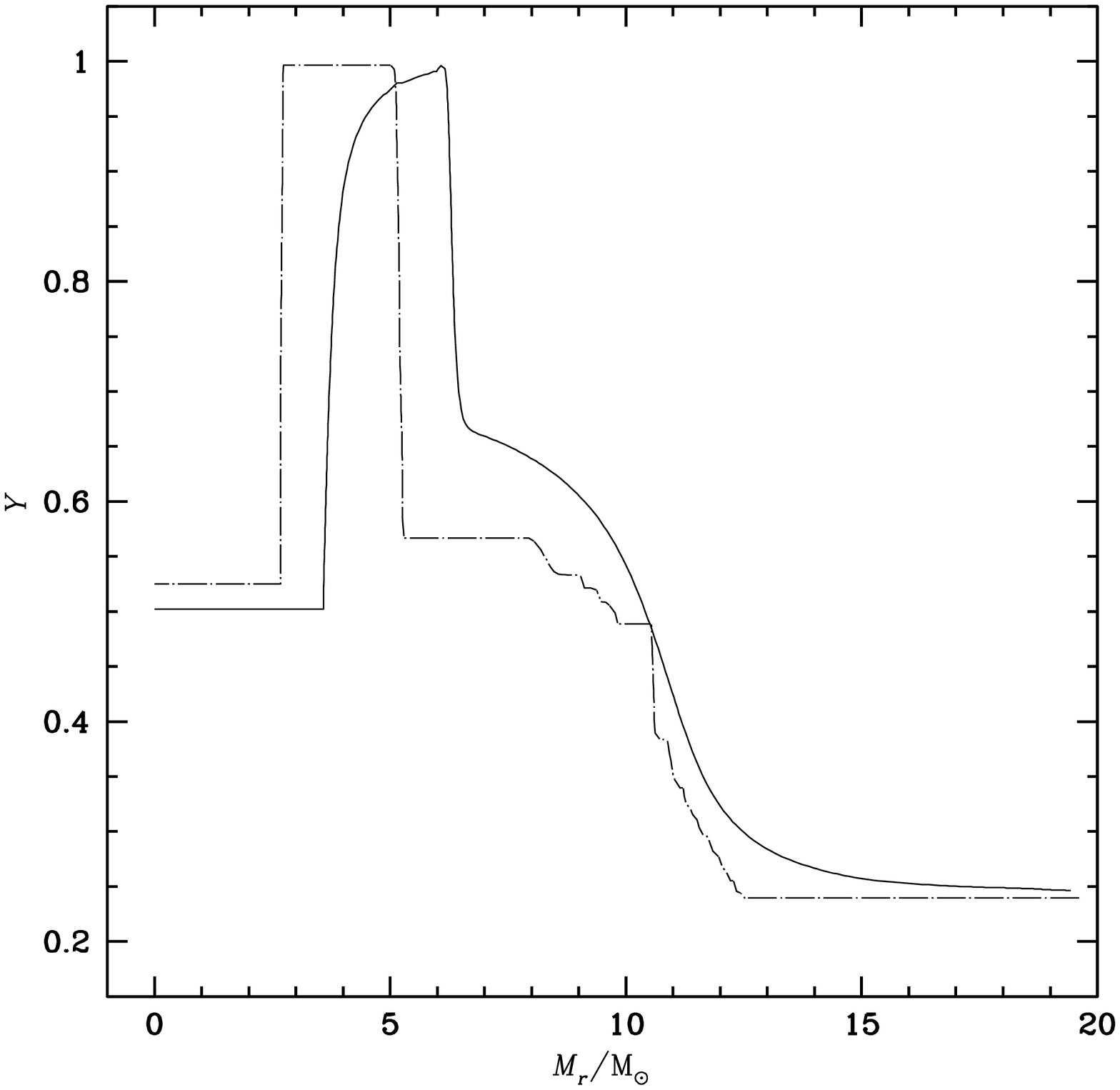}
\caption{{\it Left panel}: Evolution of the $T_{\mathrm{eff}}$
as a function of the fraction of the lifetime spent
in the He--burning phase for 15, 20 and 25
 M$_\odot$ stars at $Z$ = 0.004 with $v_{\mathrm{ini}}$ = 0
and 300 km s$^{-1}$.  
{\it Right panel}: Comparison of the internal distribution of helium
in two models of 20 M$_\odot$ at the middle of the
He--burning phase. The dashed--dot line concerns the models
with zero rotation and the continuous line represents
the case with $v_{\mathrm{ini}}$ = 300 km s$^{-1}$.}\label{fig2}
\end{figure}

Fig.~\ref{fig2}, {\it left panel}, shows for 
the models of  15, 20 and 25 M$_{\odot}$
 the changes of $T_{\mathrm{eff}}$
as a function of the fractional lifetimes in the He--burning
phase for different  rotation. For all masses, we notice that
 the non--rotating stars spend nearly the whole of 
their He--phase  as blue supergiants  and almost none as 
red supergiants. For $v_{\mathrm{ini}}$ = 300 km s$^{-1}$ (which corresponds
to about $\overline{v}$ = 220 km s$^{-1}$), 
we notice a drastic
decrease of the blue phase and a corresponding large 
increase of the red supergiant phase. This figures illustrates the fact that
models including the effects of rotation evolves much more rapidly into the red supergiant phase than non-rotating models.
Of course we have not yet the complete answer to the problem of the B/R ratio. For this we should show that we can reproduce the plots shown in Fig.~\ref{fig1}, but still part of the problem
appears to be resolved by rotation.

The physical reason for this behavior is the following:
first, let us recall that, for a given luminosity,
an extended convective zone allows
the star to remain more compact, the gradient of density being shallower in a convective zone than
in a radiative one. In the rotating model, the 
convective zone associated to the H-burning shell rapidly disappears.
This is mainly caused by the
mild mixing just outside the core produced 
by rotation during the MS phase. This mixing increases 
the amount of helium near and above the H--shell, and decreases the abundance of hydrogen there.
This makes the H-burning shell less active
(see for more details Maeder \& Meynet 2001). As a consequence the convective zone associated to it
rapidly disappears making the star to evolve into the red supergiant stage.
This is illustrated in Fig.~\ref{fig2}, {\it right panel},
which shows models of a 20 M$_\odot$ star in the middle
of the He--burning phase. One sees that: 1) the mixing in the MS phase leads  to a
slight  extension of the core (this also  
favors a redwards motion during the He--burning phase); 2) the higher He abundance
in the region of the H-burning shell (the H--shell
is just on the right side of the big He--peak) which makes this region radiative (there
is an abundance gradient) instead of being convective as in the non-rotating model (flat profile).

\section{The Wolf-Rayet stars}

Wolf--Rayet stars play a very important role in Astrophysics, as signatures
of star formation in galaxies and starbursts, as injectors of chemical elements and of the 
radioactive isotope $^{26}$Al, as  sources of kinetic energy into the interstellar medium and 
as progenitors of supernovae and, likely, as progenitors of long soft $\gamma$--ray bursts.

Let us recall
some difficulties faced by standard stellar models concerning the WR stars.
  A good agreement between 
the predictions of the stellar models for the WR/O number ratios and the observed 
ones at different metallicities in regions of constant star formation was achieved 
provided the mass loss rates were enhanced by about a factor of two during the MS
and WNL phases (Maeder \& Meynet 1994). This solution, which at that time appeared 
reasonable in view of the uncertainties pertaining the mass loss rates, is no longer
applicable at present. Indeed, the mass loss rates during the WR phase are
reduced by a factor 2 to 3, when account is given to the clumping effects
in the wind (Nugis and Lamers 2000).
Also, the mass loss rates for O--type stars have been substantially revised 
(and in general reduced) by the new results of 
Vink et al. (2001).
In this new context, it is quite clear that with these new mass loss rates
the predicted numbers of WR stars by standard non--rotating models would be much
too low with respect to the observations.

A second difficulty of the standard models with mass loss concerns
the observed number of transition WN/WC stars. These stars show simultaneously some 
nitrogen characteristic of WN stars and some carbon of the further WC stage.
The observed frequency of WN/WC stars among WR stars turns around 4.4 \% (van der Hucht 2001), while
the frequency predicted by the standard models without extra--mixing processes 
are lower by 1--2 orders of magnitude (Maeder \& Meynet 1994).
A third difficulty of the standard models as far as WR stars were concerned was that
there were relatively too many WC stars with respect to WN stars predicted (see the review
by Massey 2003). These difficulties are the signs that some process is missing
in standard models. 

Models with rotation can account for the observed variation of the number of WR to O-type stars
with the metallicity, while non-rotating models predict much too low values for these ratios
when loss rates accounting for clumping effect are used (Meynet \& Maeder 2005).

\begin{figure}
\includegraphics[width=2.5in,height=3in]{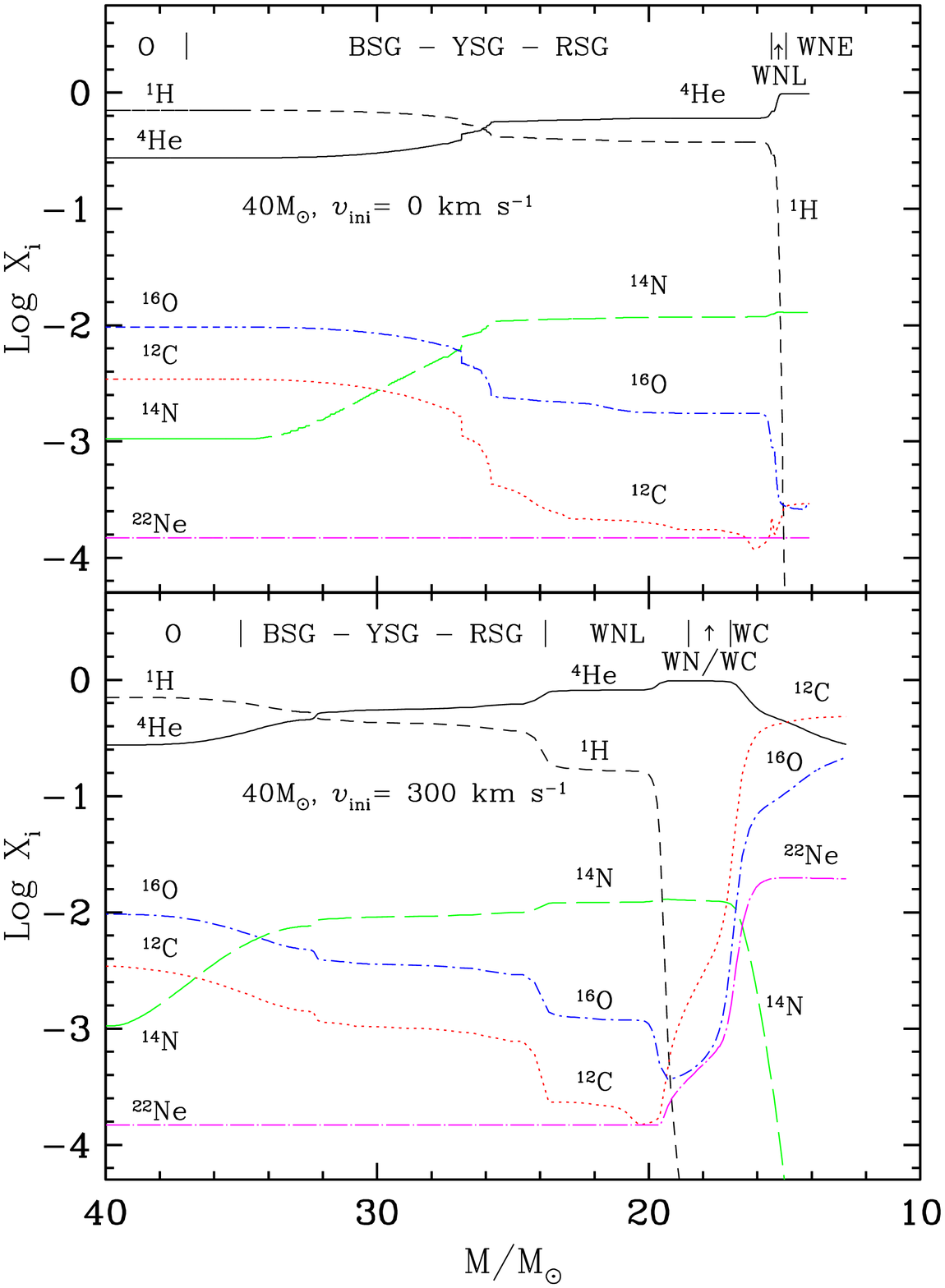}
\hfill
\includegraphics[width=2.5in,height=2.1in]{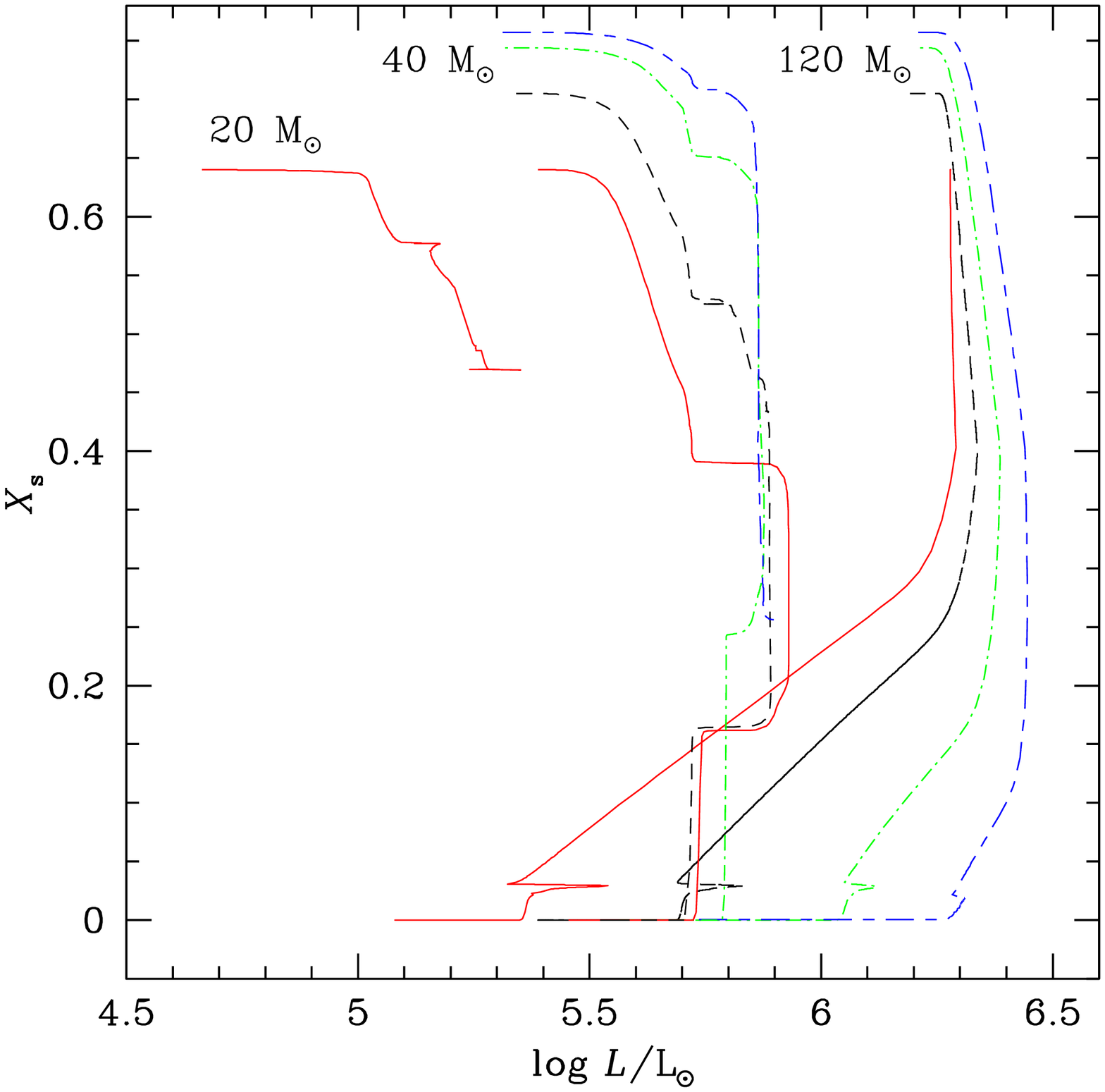}
\caption{{\it Left panel}: Evolution as a function of the actual mass of the 
abundances (in mass fraction) at the surface of a non--rotating  
(upper panel) and a rotating (lower panel) 40 M$_\odot$ stellar model.  
{\it Right panel}: Evolutionary tracks in the $X_{\rm s}$ versus log {\it L}/ {\it L}$_\odot$ plane,
where $X_{\rm s}$ is the hydrogen mass fraction at the surface. The initial masses are indicated.
Long--short dashed curves show the evolution of $Z$ = 0.004 models, 
dashed--dotted curves, short--dashed curves
and continuous lines show the evolutions for $Z$ = 0.008, 0.020 and 0.040 respectively.}\label{fig3}
\end{figure}

Fig.~\ref{fig3} left shows the evolution of surface abundances for a 40 $M_{\odot}$ star 
with and without rotation. Rotation makes  smoother changes of abundances, due to internal mixing.
As a consequence more stars are predicted to be in the intermediate WN/WC phase as required
by the observations.


Fig.~\ref{fig3} right shows the evolution of the H--surface content $X_{\mathrm{s}}$ vs. luminosity.
This is a very constraining diagram especially for the transition stages from O, Of, LBV to
WN stages. It is useful for establishing the proper filiations between  such stars. It clearly 
supports the view that in general WN stars succeeds the LBV stage (although
there may be back and forth evolution between LBV and WNL stars).
We also see that descendants from high masses at higher $Z$ significantly decrease
in luminosity during their  evolution.

The rare WO stars, characterized by a high O/C ratio represent a more advanced stage of nuclear
processing. Curiously enough, such stars which may be the progenitors of supernovae SNIb/c
are found  only at lower $Z$. The physical reasons of that have been explained by Smith \& Maeder (1991):
lower $Z$ stars become WC stars (if they do it) only very late in evolution, 
i.e.~with a high O/C ratio. 
At the opposite, at higher $Z$ the WC stars may occur at an early stage of He--processing, i.e.
with a low O/C ratio. 
This is noticeable  because of the possible
connection WO stars - SNIb/c - GRBs ($\gamma$ Ray Bursts). In this context,
we consider WO stars as good candidate for GRB progenitors (Hirschi et al. 2005; Meynet \& Maeder 2007).

\section{Primary nitrogen production by massive stars needed at low Z}

The observation of the N/O abundance ratio at the surface of metal poor halo star (see Spite et al. 2005) has put very 
interesting constraints on the nitrogen nucleosynthesis at low Z. To reproduce these observations
it is necessary that massive stars produce large amount of primary nitrogen (Chiappini et al. 2005). Standard models
cannot do the job unless some ad hoc extra mixing process is included in the stellar models.
Thus again here, standard models show their limit. 

Low metallicity, rotating massive stars naturally produce
primary nitrogen in quantities which increase with the initial velocity (Meynet \& Maeder 2002). In order
to reproduce the observations, initial rotation velocities of the order of 50\% the critical velocity
has to be assumed (Chiappini et al. 2006). Interestingly this may not only account for the
production of primary nitrogen, but also for the upturn of the C/O ratio observed at low $Z$ and
for the scatter in the N/O ratios. Such rotations may considerably change our picture of the evolution
massive stars at very low $Z$
because these stars would undergo important mass loss through various mechanisms
induced by their fast rotation: many consequences are expected bearing on the origin of the
C-rich extremely metal poor stars (Meynet et al. 2006), of the He-rich stars in Omega Cen (Maeder \& Meynet 2005) and of the anti correlations
in the abundances of light elements at the surface of globular cluster stars (Decressin et al. 2007).

To conclude, rotation appears as a very important physical ingredients of stellar models. New grids
of stellar models accounting for the effects of rotation, spanning the whole mass range of stars and all evolutionary stages are now in preparation for various metallicities.
This new sets of data will hopefully provide a new tool for studying the stellar populations in
galaxies.







\end{document}